\documentclass[prl,twocolumn,superscriptaddress,aps,longbibliography,showpacs,preprintnumbers,amsmath,amssymb,floatfix]{revtex4-1}
\usepackage{titlesec}
\usepackage{amsmath}
\usepackage{bm}
\usepackage{graphicx}
\usepackage[ansinew]{inputenc}
\usepackage{array}
\usepackage{color}
\usepackage{amsxtra}
\usepackage{amstext}
\usepackage{amssymb}
\usepackage{latexsym}
\usepackage{gensymb}
\usepackage{dsfont}
\usepackage{braket}
\usepackage[caption=false]{subfig}
\usepackage[makeroom]{cancel}

\usepackage{balance}

\usepackage{dcolumn}
\usepackage{bm}
\usepackage{bbm}
\usepackage{braket}
\usepackage{mathrsfs}
\usepackage{amstext}
\usepackage{euscript}
\usepackage{txfonts}

\usepackage[unicode=true,
 bookmarks=false,
 breaklinks=false,pdfborder={0 0 1},backref=false,colorlinks=true,citecolor=blue,urlcolor=blue,linkcolor=blue]
 {hyperref}

\def\Tr{\text{Tr}}

\def\bs{\boldsymbol}

\def\M{\mathcal{M}}

\def\I{\mathcal{I}}
\def\wt{\tilde}
\def\CM{\bs{\gamma}}
\def\P{\mathcal{P}}

\makeatother

\newcommand{\cref}[1]{Ref.\,\cite{#1}}

\newcommand*{\figref}[2][]{%
  \hyperref[{#2}]{%
    \ref*{#2}%
    \ifx\\#1\\%
    \else
      #1%
    \fi
  }%
}

\makeatother
\makeatletter
\renewcommand{\p@subsection}{}
\renewcommand{\p@subsubsection}{}
\makeatother

\begin{document}

\author{Jiru Liu}
\email{ljr1996@tamu.edu}
\affiliation{Institute for Quantum Science and Engineering (IQSE) and Department of Physics and Astronomy, Texas A\&M University, College Station, TX 77843-4242, USA}

\author{Wenchao Ge}
\email{wenchao.ge@uri.edu}
\affiliation{Department of Physics, University of Rhode Island, Kingston, Rhode Island 02881, USA}

\author{M. Suhail Zubairy}
\email{zubairy@physics.tamu.edu}
\affiliation{Institute for Quantum Science and Engineering (IQSE) and Department of Physics and Astronomy, Texas A\&M University, College Station, TX 77843-4242, USA}
\date{\today}

\title{Classical-Nonclassical Polarity of Gaussian States}

\begin{abstract}
Gaussian states with nonclassical properties such as squeezing and entanglement serve as crucial resources for quantum information processing. Accurately quantifying these properties within multi-mode Gaussian states has posed some challenges. To address this, we introduce a unified quantification: the 'classical-nonclassical polarity', represented by $\mathcal{P}$. For a single mode, a positive value of $\mathcal{P}$ captures the reduced minimum quadrature uncertainty below the vacuum noise, while a negative value represents an enlarged uncertainty due to classical mixtures. For multi-mode systems, a positive 
$\mathcal{P}$ indicates bipartite quantum entanglement. We show that the sum of the total classical-nonclassical polarity is conserved under arbitrary linear optical transformations for any two-mode and three-mode Gaussian states. For any pure multi-mode Gaussian state, the total classical-nonclassical polarity equals the sum of the mean photon number from single-mode squeezing and two-mode squeezing. Our results provide a new perspective on the quantitative relation between single-mode nonclassicality and entanglement, which may find applications in a unified resource theory of nonclassical features.
\end{abstract}
\maketitle


Gaussian states are continuous-variable quantum systems that are not only straightforward to describe from a theoretical standpoint, but also convenient to
produce and manipulate experimentally \cite{koga2012dissipation,reck1994experimental}. Nonclassical Gaussian states such as single-mode squeezed vacuum states and the Einstein-Podolski-Rosen state are essential resources for quantum-enhanced applications~\cite{lloyd1999quantum,braunstein2005quantum,wang2007quantum, weedbrook2012gaussian, adesso2014continuous}, including  
quantum teleportation~\cite{van2000multipartite}, quantum dense coding~\cite{braunstein2000dense}, quantum computing \cite{ menicucci2008one, lund2014boson}, and quantum sensing \cite{sparaciari2016gaussian,nichols2018multiparameter,zhuang2018distributed,liu2020quantum}. The nonclassical features of multimode Gaussian states could arise from a reduced quadrature variance of a single mode below the vacuum noise or/and quantum entanglement between two or multiple modes. In particular, a single-mode squeezed vacuum state can be distributed in a multimode linear optical network of beam splitters (BSs) and phase shifters to create multipartite entangled states~\cite{van2000multipartite, Kim2002, armstrong2012programmable}. In general, qualitative aspects of nonclassicality conversion have been explored~\cite{vogel2014unified, killoran2016converting}. However, a quantitative understanding of these nonclassical properties in multimode Gaussian states is crucial for evaluating the enhancement in quantum information applications.

Various quantifications have been proposed to evaluate the degree of single-mode nonclassicality~\cite{lee1991measure, asboth2005computable, marian2002quantifying, gehrke2012quantification, ge2020evaluating} and bipartite entanglement~\cite{werner2001bound,adesso2004extremal,adesso2004quantification,wang2007quantum,giedke2001quantum,vidal2002computable} individually. 
For a single-mode state, the Lee nonclassicality depth quantifies the minimum number of thermal photons necessary to destroy whatever nonclassical effects exist in the quantum state~\cite{lee1991measure}. Resource theories of single-mode quantum states~\cite{tan2017quantifying, yadin2018operational, kwon2019nonclassicality, geOpertional2020} have been explored to determine their usefulness as a resource, e.g, in metrology~\cite{geOpertional2020}. For two-mode Gaussian states, the entanglement of formation gives the amount of the entropy of the state minimized from all possible state decomposition~\cite{giedke2003entanglement}. Yet, it is difficult to calculate this quantity in general~\cite{marian2008entanglement}. An easy entanglement measure to compute is the logarithmic negativity, which quantifies how much the state fails to satisfy the positive partial transpose (PPT) condition~\cite{simon2000peres,plenio2005logarithmic}. The logarithmic negativity can be written as an analytical function of the minimum symplectic eigenvalue of the partially-transposed state and it can quantify the degree of bipartite entanglement of a $1\times (n-1)$ modes Gaussian state~\cite{serafini2006multimode}. Yet, it remains an elusive task to define a unified quantification for both single-mode and multi-mode nonclassicalities. Furthermore, exploring the conversion between these nonclassicalities adds another layer of complexity.

\par
\begin{figure}
\begin{center}
\includegraphics[width=8.5cm]{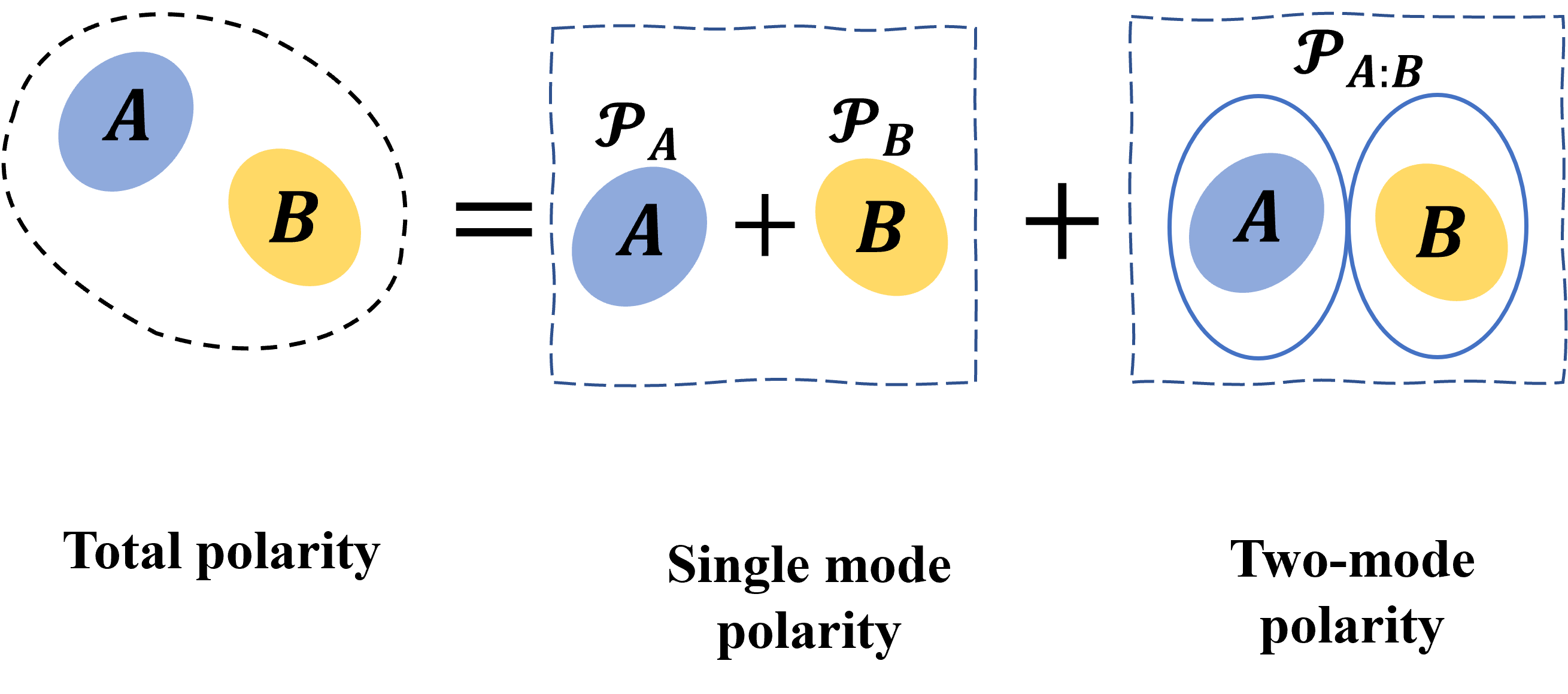}
\caption{Total classical-nonclassical polarity of a two-mode Gaussian state.}
\label{fig:two}
\end{center}
\end{figure}

Several previous works  \cite{ge2015conservation,liu2021distribution,arkhipov2016nonclassicality,tahira2009entanglement,hertz2020relating} have discussed the quantitative conversion of nonclassicality and entanglement during BS operations. For example, Ge et al. have explored a conservation relation of the two quantities during BS transformation for certain two-mode Gaussian states \cite{ge2015conservation}. Moreover, Arkhipov et al. have found an invariant for nonclassical two-mode Gaussian states which comprises the terms describing both
local nonclassicality of the reduced states and the entanglement of the whole system related to the symplectic eigenvalues \cite{arkhipov2016nonclassicality}, and extended the results to pure three-mode Gaussian states. However, 
these results only hold for a subset of two-mode or three-mode Gaussian states.

\begin{figure}
\begin{center}
\includegraphics[width=8.5cm]{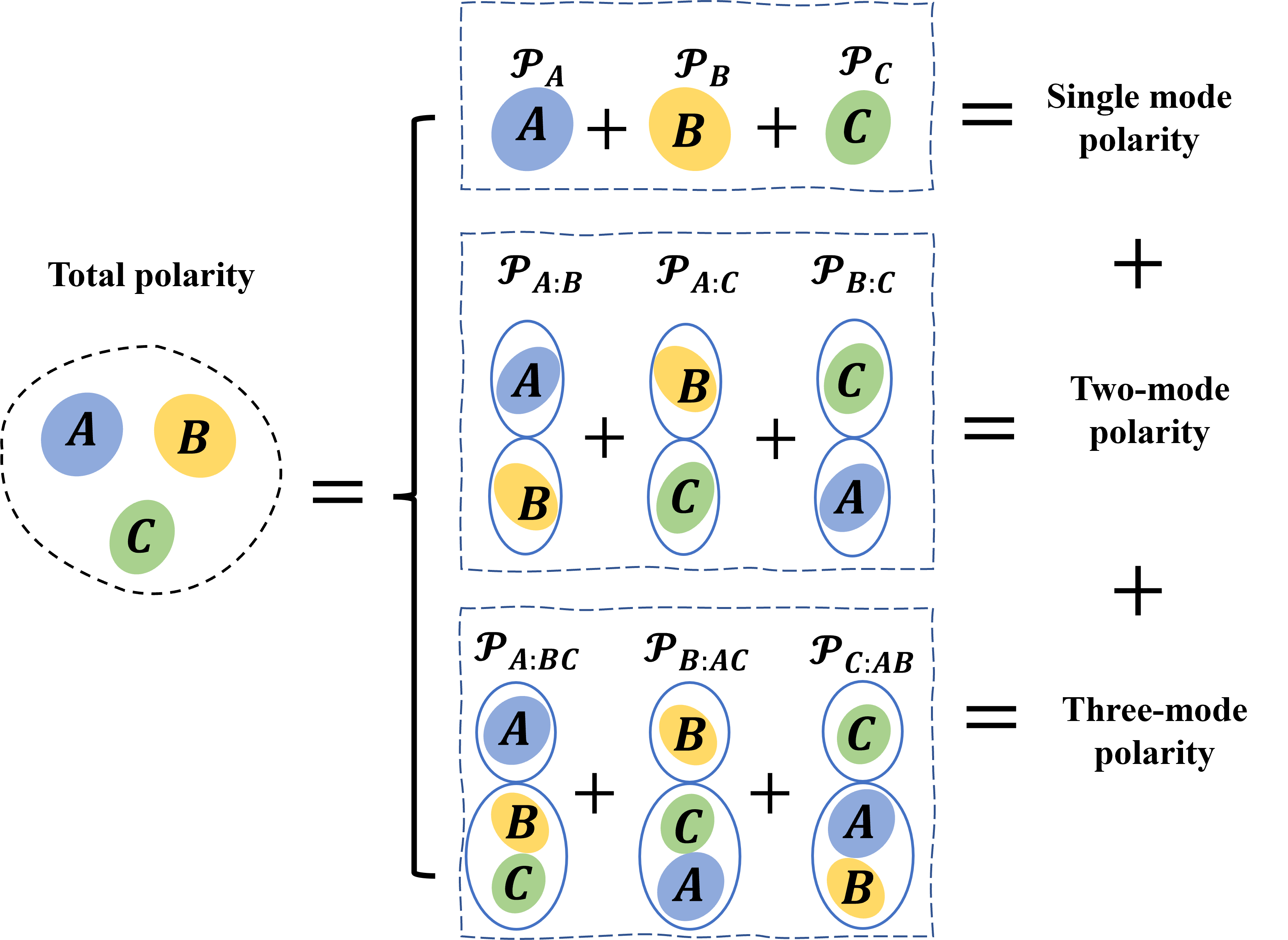}
\caption{Total classical-nonclassical polarity of a three-mode Gaussian state. }
\label{fig:three}
\end{center}
\end{figure}

In this Letter, we aim to establish a unified quantification for both single-mode nonclassicality and multimode bipartite entanglement that is invariant under passive linear transformations. We introduce classical-nonclassical polarity (CNP) for quantifying single-mode nonclassicality and $1\times(n-1)$-mode bipartite entanglement. Contrary to the measure quantification of quantum resource theories~\cite{chitambar2019quantum,streltsov2017colloquium}, our definitions can be positive or negative. This dual nature is necessitated by the conservation relation to hold for both classical and nonclassical Gaussian states. By taking into account the CNPs of the reduced single modes and all possible bipartite modes, we find the total CNP of an arbitrary two-mode or three-mode Gaussian state is a linear function of the symplectic invariants~\cite{serafini2006multimode} and the extreme quadrature variances of the reduced single-mode states. 
Moreover, we show that the total CNP is a conserved quantity for arbitrary two-mode and three-mode Gaussian states under an optical linear network (Fig. \ref{fig:two} and \ref{fig:three}). Our results provide a new perspective on the complex structure of nonclassical features in multi-mode Gaussian states, which may find applications in a unified resource theory of nonclassical states in different physical settings.


%

\textit{Gaussian state preliminary--}An $n$-mode Gaussian state, whose density matrix is denoted as $\rho$, is fully characterized, up to local displacements, by its covariance matrix (CM) $\bs{\gamma}$ of elements $ \bs{\gamma}_{jk} = \frac 12\braket{\Delta \hat{\bs{r}}^\dag_j\Delta\hat{\bs{r}}_k+\Delta\hat{\bs{r}}_k\Delta\hat{\bs{r}}^\dag_j}$, where  $\braket{\hat{X}}=\Tr(\rho\hat{X})$, $\Delta\hat{\bs{r}}_j$ denotes $\hat{\bs{r}_j}-\braket{\hat{\bs{r}}}_j$, and $\hat{\bs{r}}=[\hat{a}_1^\dag,\hat{a}_1,...\hat{a}_n^\dag,\hat{a}_n]^\dag$ is the vector of the bosonic field operators \cite{giedke2001quantum,supple}. For a two-mode Gaussian state $\rho_{AB}$, the CM $\bs{\gamma_{AB}}$ is given by a $4\times 4$ matrix
\begin{equation}\label{eq:gamma_2}
    \bs{\gamma_{AB}} = \left[\begin{matrix}
        \bs{\gamma_A} & \bs{x}^\dag \\
        \bs{x} & \bs{\gamma_B}
    \end{matrix}
    \right].
\end{equation}
Note that $\bs{\gamma_A}$ ($\bs{\gamma_B}$) is a $2\times 2$ matrix, representing the reduced single-mode Gaussian state $\rho_A$ ($\rho_B$), and $\bs{x}$ describes the correlation of the two modes. 

\
\textit{Symplectic invariants in Gaussian states--} Gaussian operations refer to unitary transformations that map a Gaussian state onto another Gaussian state. The exponent of these unitaries consists of terms up to quadratic in the bosonic field operators \cite{weedbrook2012gaussian}.
When a Gaussian state $\rho$ undergoes a Gaussian unitary transformation $\hat{U}$, it induces a symplectic transformation $\bs{S}$ on its associated covariance matrix $\CM$. This correspondence can be succinctly expressed as: $ \rho'= \hat{U}^\dag\rho\hat{U} \rightleftharpoons \bs{\gamma'} = \bs{S\gamma S}^\dag$ \cite{simon1994quantum}.
According to Williamson's theorem \cite{arnol1989mathematical}, the CM $\bs{\gamma}$ can always be symplectic diagonalized, namely, $\bs{S\gamma S^\dag} = \text{diag}(\nu_1,\nu_1,\nu_2,\nu_2,\cdots, \nu_n, \nu_n)$. Here $\nu_j$ are the symplectic eigenvalues. Certain quantities remain invariant under symplectic transformations, which are defined as symplectic invariants. As discussed in Ref.~\cite{serafini2006multimode,serafini2023quantum}, a natural choice of symplectic invariants for $n$-mode Gaussian states is given by \cite{serafini2006multimode}
\begin{equation}
    \I^{(n)}_k = \sum\M_{n-k}(\CM),
\end{equation}
where $\M_{n-k}(\CM)$ represents the minors of order $(n-k)$, which are obtained by calculating the determinants of sub-matrices that result from removing $k$ rows and $k$ columns from the block matrix. Given that there are various ways to select which rows and columns to delete, multiple minors of the same order can be derived. The summation encompasses all possible minors of the $(n-k)-$th order \cite{supple}. In terms of the symplectic diagonalized form of the CM, we have \cite{adesso2006continuous}
\begin{align}\label{eq:Vieta}
\I_{k}^{(n)}=\sum_{\mathcal{S}_k^n}\prod_{j\in \mathcal{S}_k^n} \nu_j^2,
\end{align}
where $\mathcal{S}_k^n$ represents a subset of $n-k$ integers chosen from integers $1,2,...,n$ and the summation goes over all possible subsets \cite{supple}.
\par
For a general scenario, consider a two-mode Gaussian state whose CM is given by Eq. (\ref{eq:gamma_2}). Two symplectic invariants can be identified: $\I_0^{(2)} =|\bs{\gamma_{AB}}|$ and $\I_1^{(2)} = |\bs{\gamma_A}|+|\bs{\gamma_B}|+2|\bs{x}|$ \cite{wang2007quantum,giedke2001quantum, Serafini_2004}. Note that $|\cdot|$ signifies the matrix determinant. 
\par
For a three-mode Gaussian state, represented as $\rho_{ABC}$, the CM is described by
\begin{equation}\label{eq:gamma3}
    \bs{\gamma_{ABC}} = \left[\begin{matrix}
        \bs{\gamma_A} & \bs{x}^\dag & \bs{z}^\dag\\
        \bs{x} & \bs{\gamma_B} & \bs{y}^\dag\\
        \bs{z} & \bs{y} & \bs{\gamma_C}
    \end{matrix}
    \right],
\end{equation}
where $\bs{x,y,z}$ denote the interaction of modes $A,B$, modes $B,C$, and modes $A,C$, respectively.
The symplectic invariants of the three-mode Gaussian state are $\I_0^{(3)}=|\bs{\gamma_{ABC}}|$, $\I_1^{(3)} =|\bs{\gamma_{AB}}|+|\bs{\gamma_{BC}}|+|\bs{\gamma_{AC}}|+2|\bs{D_x}|+2|\bs{D_y}|+2|\bs{D_z}|$, and $\I_2^{(3)} =|\bs{\gamma_{A}}|+|\bs{\gamma_{B}}|+|\bs{\gamma_{C}}|+2|\bs{x}|+2|\bs{y}|+2|\bs{z}|$, where $\bs{D_x},\bs{D_y},\bs{D_z}$ are $4 \times 4$ matrices given by:
$    \bs{D_x} = \left[\begin{matrix}
        \bs{x} & \bs{\gamma_B}\\
        \bs{z} & \bs{y} 
    \end{matrix}
    \right],\quad\bs{D_y} = \left[\begin{matrix}
        \bs{y} & \bs{\gamma_C}\\
        \bs{x}^\dag & \bs{z}^\dag 
    \end{matrix}
    \right],\quad\bs{D_z} = \left[\begin{matrix}
        \bs{z} & \bs{y}\\
        \bs{\gamma_A} & \bs{x}^\dag 
    \end{matrix}
    \right].$ 
We explain the patterns of the minors for a three-mode system $\bs{\gamma_{ABC}}$ in \cite{supple}. 
\par 

\textit{Single-mode classical-nonclassical polarity--} In order to quantify the nonclassicality of each single mode (tracing out the rest modes~\cite{ge2015conservation, arkhipov2016nonclassicality}) in an $n$-mode Gaussian state, we introduce the concept of single-mode classical-nonclassical polarity, denoted as $\P^{(1)}$. Unlike the definition of a nonclassicality measure \cite{chitambar2019quantum}, which is strictly positive-definite, a negative CNP value means that a state is more classical than coherent states.
For a single-mode Gaussian state with the CM $ \bs{\gamma}^{(1)}=\left[\begin{matrix}
        a & b^* \\
        b & a
    \end{matrix}
    \right]$, the degree of polarity is defined as
\begin{equation}\label{eq:single_CNP}
    \P^{(1)} = -(\lambda-\frac 12)(\Lambda - \frac 12),
\end{equation}
where $\lambda = a-|b|, \Lambda = a+|b|$ are the minimum and the maximum eigenvalues of the CM matrix $\bs{\gamma}^{(1)}$, describing the minimum and the maximum quadrature variances of the quantum state. In the context of single-mode Gaussian states, $|\bs{\gamma}|=\lambda\Lambda\geqslant 1/4$ \cite{wang2007quantum}. Thus, $\Lambda\geqslant1/2$ always holds. For $\lambda<1/2$, the state has the minimum quadrature variance below that of coherent states \cite{supple}, therefore it is nonclassical and the CNP $\P^{(1)}>0$. For $\lambda>1/2$, the state has the minimum quadrature variance larger than that of coherent states, meaning classical and $\P^{(1)}<0$. When $\lambda=1/2$, $\P^{(1)}=0$ is the classical-nonclassical boundary, where the states are squeezed thermal coherent states with the minimum quadrature variance being $1/2$. For a pure state, it can be calculated that $\P^{(1)}=\braket{\hat a^{\dagger}\hat a}-|\braket{\hat a}|^2$, which quantifies the amount of averaged photon number from the nonclassical process, i.e., single-mode squeezing.
\par
Our definition of single-mode CNP takes the purity $\left(4|\bs{\gamma}^{(1)}|\right)^{-1}$ \cite{wang2007quantum} of the state into account. The smaller the purity, the greater the absolute value of $\P^{(1)}$. An arbitrary single-mode Gaussian state can be described as a squeezed thermal coherent state~\cite{weedbrook2012gaussian}, where $\lambda=(1/2+\braket{n_{\text{th}}})e^{-2r}$ and $\Lambda=(1/2+\braket{n_{\text{th}}})e^{2r}$ \cite{supple}. Here $r$ is the squeezing parameter and $\braket{n_{\text{th}}}$ is the averaged number of thermal photons. For nonclassical states with the same value of $\lambda$, smaller purity is represented by a larger value of $\Lambda$. Thus, the CNP for nonclassical states characterizes the degree of squeezing for both pure and mixed states.

\emph{$1\times (n-1)$ modes bipartite classical nonclassicality polarity--}For a single-mode Gaussian state, the CNP characterizes the squeezing property, while in a multi-mode scenario, it is the entanglement between different subsystems that introduces another degree of nonclassicality. For $1\times (n-1)$ mode Gaussian states, the positive partial transpose (PPT) criterion \cite{werner2001bound,wang2007quantum,simon2000peres} is both necessary and sufficient. Therefore, violation of the criterion can be used to construct bipartite entanglement measures. For a partial-transposed state $\rho^{T_A}$ (without loss of generality, transpose the first mode $A$ here), the corresponding CM is $\bs{\gamma^{T_A}}=\bs{T\gamma T}$ with $\bs{T}=\bs{t}\oplus \bs{\mathbbm{1}}_{2n-2}$, where $\bs{t}=\left[\begin{matrix}
        0 & 1 \\
        1 & 0
    \end{matrix}
\right]$ and $\bs{\mathbbm{1}}_{2n-2}$ is the $(2n-2)$-dimension identity matrix \cite{simon2000peres}. For convenience, denote $\bs{\gamma^{T_A}}$ as $\wt{\CM}$. The symplectic invariants $\wt{\I}^{(n)}_{j}$ of the PPT state $\rho^{T_A}$ are the summations of minors of $\wt{\CM}$, which can be expressed by the symplectic eigenvalues $\tilde{\nu}_j$ of $\tilde{\bs{\gamma}}$ in the same form as Eq. (\ref{eq:Vieta}).
According to the PPT criterion, if $\wt{\nu}_j\geqslant 1/2$, $\rho$ is a PPT state, implying mode $A$ is separable from other subsystems. Ref.~\cite{serafini2006multimode} demonstrates that at most one symplectic eigenvalue can be smaller than $1/2$. Therefore, computable entanglement measures have been defined by comparing the minimum partially-transposed symplectic eigenvalue $\tilde{\nu}_{\min}$ with $1/2$, such as the negativity and the logarithmic negativity, both of which quantify how much the PPT condition is violated. However, these quantifications do not differentiate states with the same minimum symplectic eigenvalue but with other different properties, such as the purity of the system. 
\par
In this context, we introduce the CNP for bipartite $1\times(n-1)$-mode Gaussian states based on the symplectic invariants of PPT state.
Regarding the PPT state, Eq. (\ref{eq:Vieta}) suggests an analogy with Vieta's formulas if we treat $\wt{v}^2_i$ as the roots of a polynomial. With this insight, we define the polynomial functions $g^{(n)}(x)\equiv 2\sum_{j=0}^n (-1)^{j+1}x^{j}\wt{\I}^{(n)}_{j}$ of degree $n$ $(n\geqslant 2)$, where $\wt{\I}_n^{(n)}$ is set to be 1. Through Vieta's formulas, the $g(x)$ function can be expressed as: $g^{(n)}(x)=-2\prod_{j=1}^n(\tilde{v}_j^2-x)$. Therefore, the separability condition $\tilde{\nu}_j\ge 1/2$ is equivalent to $g^{(n)}(1/4)\le 0$. It reveals that $g^{(2)}(1/4)> 0$ indicates two-mode Gaussian entanglement exists, while $g^{(n)}(1/4)< 0$ leads to separable states. Hence we define the $1\times(n-1)$ modes bipartite CNP as 
\begin{equation}\label{eq:multi_CNP}
\P^{(n)}=g^{(n)}(\frac 14)=-2\prod_{j=0}^n\left(\tilde{v}^2_j-\frac{1}{4}\right).
\end{equation}
\par
Taking $n=2$ for example, the sign of $\P^{(2)}$ determines whether the two-mode bipartite system is entangled or separable, while the absolute value quantifies the distance of the state to the separability-entanglement boundary. Higher-order of the multimode CNPs quantifies any additional separability or entanglement contribution to the $1\times (n-1)$ modes bipartite system. In terms of the symplectic invariants, the two-mode CNP is given by
\begin{align}
\label{eq:xi}
\P^{(2)}_{A:B} 
    &=-\frac 1{8}+\frac 12 \wt{\I}^{(2)}_1-2\wt{\I}^{(2)}_0.
\end{align}
In \cite{supple}, we show that $\wt{\I}^{(2)}_1=\I^{(2)}_1-4|\bs{x}|=|\bs{\gamma_A}|+|\bs{\gamma_B}|-2|\bs{x}|$ and $\wt{\I}^{(2)}_0=\I^{(2)}_0=|\bs{\gamma_{AB}}|$. Similar to the single-mode CNP, the two-mode CNP exhibits linearity in relation to (sub-)matrix determinants and is influenced by the system's purity. For entangled states, the value of $\P^{(2)}$ takes into account the potential extra effort to prepare a more mixed state with the same $\tilde{v}_{\min}$. For separable states, a smaller purity means more classical.
\par
In a three-mode system, denoted as $\rho_{ABC}$, bipartite CNP can manifest in two ways: as two-mode polarity and as three-mode polarity, as illustrated in Fig. \ref{fig:three}. For the two-mode CNP, by tracing out one mode from $\rho_{ABC}$, it encompasses three components: $\P^{(2)}_{A:B}$, $\P^{(2)}_{B:C}$, and $\P^{(2)}_{C:A}$. On the other hand, the three-mode CNP is divided into the following three terms: $\P^{(3)}_{A:BC}$, $\P^{(3)}_{B:CA}$, and $\P^{(3)}_{C:AB}$. These terms can indicate the bipartite entanglement or separability of the system. As an example, according to Eq. (\ref{eq:multi_CNP}), the CNP for $A:BC$ is given by
\begin{equation}\label{eq:w}
\begin{aligned}
    \P^{(3)}_{A:BC}=&\frac 1{32}-\frac {1}{8}\wt{\I}^{(3)}_2+\frac 12\wt{\I}^{(3)}_1-2\wt{\I}^{(3)}_0,
\end{aligned}
\end{equation} 
where $\wt{\I}^{(3)}_1=\I^{(3)}_1-4|\bs{D_x}|-4|\bs{D_y}|$, and $\wt{\I}^{(3)}_2=\I^{(3)}_2-4|\bs{x}|-4|\bs{z}|$ \cite{supple}. 
\par
In particular, for pure state, the following theorem holds.
\par
\textbf{Theorem 1:} \textit{For any pure multimode Gaussian state, the order of $n$ ($n\ge3$) CNP bipartite $\P^{(n)}$ equals zero.}
\par
As the two-mode polarities $\P^{(2)}_{A:B}$, $\P^{(2)}_{A:C}$ already quantify some amount of entanglement or separability of the three-mode system when one mode is traced out, the three-mode polarity $\P^{(3)}_{A:BC}$ can be zero even if the bipartite $A:BC$ is entangled. In this case, the three-mode polarity avoids over-counting the entanglement or separability of the system. For example, for a one-mode biseparable state ${\rho}_{AB}\otimes{\rho}_C$ with ${\rho}_{AB}$ entangled, $\P^{(3)}_{A:BC}=0$ when ${\rho}_C$ is pure and $\P^{(3)}_{A:BC}>0$ when ${\rho}_C$ is a mixed state. Yet, we show numerically that by adding a separable state ${\rho}_C$ to the two-mode ${\rho}_{AB}$, the total bipartite polarities in general do not increase.

\begin{figure}
\begin{center}
\includegraphics[width=8.5cm]{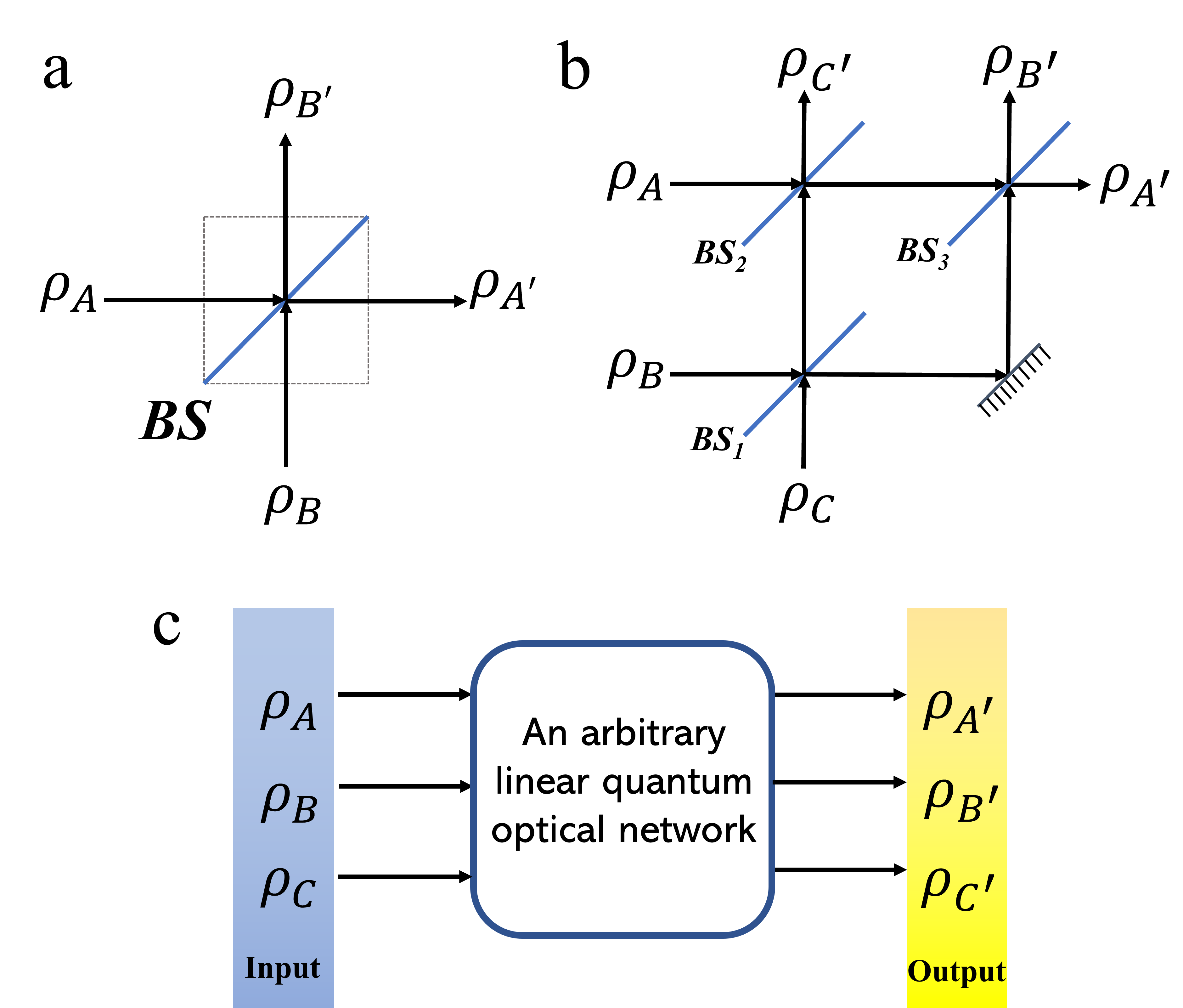}
\caption{\textbf{(a)} Two-mode Gaussian state, $\rho_{AB}$, undergoes mixing by a BS to produce a new two-mode Gaussian state, $\rho_{A'B'}$. \textbf{(b,c)} Three-mode Gaussian state, $\rho_{ABC}$, passes through networks of linear optical networks.}
\label{fig:BSnetworks}
\end{center}
\end{figure}
\emph{Conservation relation for two-mode Gaussian states before and after a beam splitter--}Beam-splitters and phase shifters are linear optical devices that do not generate additional nonclassicality \cite{Kok2017, geOpertional2020}. Entanglement can be generated from single-mode nonclassical states using beam splitters~\cite{Kim2002}. Previous work \cite{ge2015conservation} attempted to find a conservation relation of the total nonclassicality from both single-mode reduced systems and the two-mode system. However, only a subset of two-mode Gaussian states has been shown to satisfy the conservation of nonclassicality under some quantifications. Here we show that the sum of single-mode CNPs and two-mode bipartite CNP is conserved before and after a beam-splitter for arbitrary input states $\rho_{AB}$ (Fig. \figref[a]{fig:BSnetworks}). We obtain that the total CNP \cite{supple}
\begin{align}
    \P&\equiv \P_A^{(1)} + \P^{(1)}_B + \P^{(2)}_{A:B}\nonumber\\
    &=\frac 12(\lambda_A+\Lambda_A+\lambda_B+\Lambda_B) -\frac 12 \I^{(2)}_1 -\frac 58 -2\I^{(2)}_0,
\end{align}
where $\lambda_{A(B)}$ and $\Lambda_{A(B)}$ represent the minimum and the maximum eigenvalues of $\bs{\gamma_{A(B)}}$. In addition to the invariants $\I^{(2)}_1$ and $\I^{(2)}_0$, it can be shown that $\lambda_A+\Lambda_A+\lambda_B+\Lambda_B$ is also invariant before and after a BS \cite{supple}. Therefore, the total CNP of a two-mode Gaussian state is conserved under linear optical transformations. It is worth noting that although $\P$ is the sum of the total classical-nonclassical polarity, $\P<0$ does not necessarily mean the system is classical. For example, a two-mode system consisting of a weak single-mode squeezed vacuum state and a large thermal state, which has a non-positive-definite $P$ function. We note that a similar nonclassicality invariant under linear unitary transformations is introduced in Ref. \cite{arkhipov2016nonclassicality}. As an example, we calculate the total CNP for a two-mode squeezed vacuum state~\cite{yurke19862}, which gives $\P=\braket{\hat a^{\dagger}_1\hat a_1+\hat a^{\dagger}_2\hat a_2}$. We conclude that the total CNP of a single-mode squeezed vacuum equals that of a two-mode squeezed vacuum for the same average number of photons.

\emph{Conservation relation for three-mode Gaussian states in linear optical networks--}
The conservation relation can be extended to arbitrary three-mode Gaussian states using the concept of the single-mode and multimode bipartite CNPs (Fig. \figref{fig:three}) in a linear optical network.
\par
The total nonclassicality $\P$, calculated by adding single-mode, two-mode, and three-mode CNPs as shown by Fig. \figref{fig:three} are given by \cite{supple}
\begin{equation}\label{eq:total_E}
\begin{aligned}
    \P & = \sum_{\alpha={A,B,C}}\P^{(1)}_\alpha +\frac 12\sum_{\alpha,\beta={A,B,C}} \P^{(2)}_{
    \alpha:\beta}+\frac 12\sum_{\alpha,\beta,\kappa={A,B,C}}\P^{(3)}_{\alpha:\beta\kappa}\\
    &=-\frac {3}{8}\I^{(3)}_2-\frac 12\I^{(3)}_1-6\I^{(3)}_0-\frac {33}{32}+\frac 12\bar{\Lambda},
\end{aligned}
\end{equation}
where $\alpha, \beta, \kappa$ take all the permutations of $A, B, C$ in the summations, the factor 1/2 in the first line is due to the commutativity of indexes.
 $\I^{(3)}_0,\I^{(3)}_2$, and $\I^{(3)}_3$ are symplectic invariants of the three-mode system, which stay conserved during any Gaussian operation, while $\bar{\Lambda}\equiv\lambda_A+\Lambda_A+\lambda_B+\Lambda_B+\lambda_C+\Lambda_C$ also stay invariant before and after a BS \cite{supple}. Hence, $\P$ is a conserved quantity during linear optical transformations. An example of a BS network is given in Fig. \figref[b]{fig:BSnetworks}. A general scenario is shown by Fig. \figref[c]{fig:BSnetworks} where a three-mode Gaussian state is input into an arbitrary linear optical network comprising BSs, wave plates, and phase shifters \cite{kok2007linear,tan2017quantifying}. The conserved total CNP describes the conversion between single-mode and multimode nonclassicalities through a passive Gaussian transformation using our definitions Eqs. (\ref{eq:single_CNP}) and (\ref{eq:multi_CNP}).

 \textbf{Theorem 2:} \textit{For any pure multimode Gaussian state, the total classical-nonclassical polarity equals the sum of the mean photon number from single-mode squeezing and two-mode squeezing} \cite{supple}.

\textit{Discussion--}We have established a quantitative relation that links the quadrature squeezing of single modes and the bipartite entanglement of Gaussian states upto three modes. Our results on quantifying classical-nonclassical polarity of Gaussian states suggest a new method for evaluating different nonclassical properties on the same footing, which has multiple implications.
\par
First, the quantitative conversion between various nonclassical properties provides the basis for preparing quantum resources from one to another. This suggests that in order to maximize the resource output of one type, e.g. entanglement, we can design unitary transformations to deplete the input resource of the other kind~\cite{fu2020squeezing}. Second, our results provide a unified quantification for single-mode, two-mode, and three-mode Gaussian states in terms of the total classicality or nonclassicality. We can compare the degree of CNP for resources from completely different processes, such as single-mode squeezed vacuum states for $SU(2)$ interferometers and two-mode squeezed vacuum states characterized by $SU(1,1)$ interferometers~\cite{yurke19862}. The unified quantification may also support a more general resource theory of quantum states with linear optical unitaries as free operations~\cite{tan2017quantifying, yadin2018operational, kwon2019nonclassicality, geOpertional2020}. Third, our work may inspire the future study of higher-mode Gaussian states. For example, how the nonclassicality of a single-mode squeezed vacuum state is distributed in a multimode linear network for distributed quantum metrology~\cite{zhuang2018distributed, ge2018distributed}. Yet, the structure of the system would be more complex as the dimension grows. Even in the three-mode Gaussian states, our invariant quantity does not include the $1\times1\times1$ tripartite entanglement, which cannot be sufficiently and necessarily detected by the PPT criterion \cite{giedke2001separability}. 
In the case of four-mode or higher-mode Gaussian states, 
bipartite entanglement of $2\times 2$, tripartite, or quadpartite entanglement may exist but can not be fully determined by the PPT criterion~\cite{werner2001bound}.

\textit{Conclusion--}
Based on the fact that linear optical devices, including beam splitters, phase shifters, and wave plates, do not generate additional nonclassicality, we proposed the existence of a unified quantification for single-mode squeezing and multimode bipartite entanglement in terms of the classical-nonclassical polarity. 
We demonstrated that the sum of the single-mode and multi-mode classical-nonclassical polarities is conserved under linear optical transformations for arbitrary two-mode and three-mode Gaussian states. These findings highlight a quantitative conversion relation between nonclassicality and entanglement in multimode systems, enriching our understanding of multi-mode entanglement phenomena and exploiting quantum resources  across varied physical contexts. 
\par

\section*{Acknowledgments}
We thank Kangle Li from HKUST for helpful discussions. This research of is supported by the project NPRP 13S-0205-200258 of the Qatar National Research Fund (QNRF). The research of W.G. is supported by NSF Award 2243591.

\bibliography{ref}
\end{document}